\documentclass[prb,twocolumn,amsmath,amssymb]{revtex4}
\usepackage{graphicx}
\usepackage{dcolumn}
\usepackage{bm}

\begin{document}

\newcommand{\up}{\uparrow}
\newcommand{\down}{\downarrow}
\newcommand{\phdagger}{\phantom{\dagger}}          
\newcommand{\aop}[1]{a^{\phdagger}_{#1}}           
\newcommand{\adop}[1]{a^{\dagger}_{#1}}            
\newcommand{\bop}[2]{b^{\phdagger}_{#1}(#2)}           
\newcommand{\bdop}[2]{b^{\dagger}_{#1}(#2)}            
\newcommand{\cop}[1]{c^{\phdagger}_{#1}}           
\newcommand{\cdop}[1]{c^{\dagger}_{#1}}
\newcommand{\dop}[2]{d^{\phdagger}_{#1}(#2)}           
\newcommand{\ddop}[2]{d^{\dagger}_{#1}(#2)}            
\newcommand{\sS}{{\mathcal{S}}}
\newcommand{\sH}{{\mathcal{H}}}
\newcommand{\mgt}{>>}
\newcommand{\mlt}{<<}
\newcommand{\half}{\frac{1}{2}}
\newcommand{\rp}{r^{\prime}}
\newcommand{\om}{\omega}
\def\sO{{\mathcal{O}}}
\def\cd{c^{\dagger}}
\newcommand{\ket}[1]{| #1 \rangle}
\newcommand{\barpsi}[1]{\bar{\Psi}_{#1}}
\newcommand{\psiop}[1]{\Psi_{#1}}

\title{Nodal-antinodal dichotomy and magic doping fractions in a stripe ordered antiferromagnet}
\author{Mats Granath}
\email{mgranath@fy.chalmers.se}
\affiliation{G\"oteborg University\\
G\"oteborg 41296 \\
Sweden
}
\date{\today}

\begin{abstract}
We study a model of a stripe ordered doped antiferromagnet consisting of coupled Hubbard ladders which can be tuned from quasi-one-dimensional to
two-dimensional. We solve for the magnetization and charge density on the ladders by Hartree-Fock theory and find a set of solutions 
with lightly doped ``spin-stripes'' which are antiferromagnetic and more heavily doped anti-phase ``charge-stripes''. 
Both the spin- and charge-stripes have electronic spectral weight near the Fermi energy but in different regions of the 
Brillouin zone; the spin-stripes in the ``nodal'' region, near $(\pi/2,\pi/2)$, and the charge-stripes in the ``antinodal'' region, near $(\pi,0)$. 
We find a striking dichotomy between nodal and antinodal states in which the nodal states are 
essentially delocalized and two-dimensional whereas the antinodal states are quasi-one-dimensional, localized on individual charge-stripes. 
When supplemented by known results for the interacting quasi one-dimensional electron gas the present work can provide a 
framework for understanding the nodal-antinodal dichotomy of the underdoped high-T$_c$ cuprates.
For bond-centered stripes we also find an  
even-odd effect of the charge periodicity on the spectral weight in the nodal region.
We speculate that this may be related to observed non-monotonous variations with doping of the low-temperature 
resistivity in La$_{2-x}$Sr$_x$CuO$_4$.      
\end{abstract}

\maketitle

One of the most intriguing aspects of the hole-doped cuprate high-temperature superconductors is the presence of nanoscale electronic inhomogeneity. 
An important question is whether the inhomogeneity is self-organized or primarily a consequence of material disorder.
In one material,   
Bi$_2$Sr$_2$CaCu$_2$O$_{8+\delta}$ (BSCCO), it has been found by tunneling spectroscopy (STM) that that the inhomogeneity is correlated with the 
dopant disorder,\cite{Davis_disorder} 
while in materials where high resolution neutron scattering experiments have been performed, primarily La$_{2-x}$Sr$_x$CuO$_4$ (LSCO) and 
YBa$_2$Cu$_3$O$_{6+x}$ (YBCO) there appears to be a universal behavior 
of the spin correlations which is most readily interpreted in terms of stripe order or fluctuations which are clearly not a 
disorder effect.\cite{Tranquada} 
It has been suggested that stripe fluctuations are universal in the underdoped cuprates but in some materials (such as BSCCO) or doping regimes  
masked by the strong disorder or quantum fluctuations such that only the local correlations are preserved.\cite{Robertson_Sachdev,Zaanen} 
The search for the elusive {\em fluctuating} or {\em quantum} stripes is an ongoing endeavor.\cite{Reznik} Clearly, whether or not stripes are a universal 
phenomenon in the cuprates is still an open question. In parallel it may also be illuminating to study properties which are not in any obvious way related to stripes
to see if these can be understood from an assumption of stripe order or stripe correlations. Particularly important in this context is to study such properties 
of the cuprates which are believed to be universal.


A feature of the underdoped cuprates which may be universal is a sharp distinction between the nature of electronic excitations in the nodal and 
antinodal regions
of the Brillouin zone, the so called ``nodal-antinodal dichotomy''.\cite{ZX_nodal} Nodal refers to the region near the node of vanishing gap of the 
$d_{x^2-y^2}$ superconducting order parameter whereas antinodal refers to the region with maximal gap. 
The dichotomy is most vividly demonstrated in angle-resolved photoemission (ARPES) experiments on the normal state 
of underdoped LSCO where a quasiparticle peak is observed along the Fermi surface in the nodal region which disappears quite abruptly in the antinodal 
region where the spectral weight is broad and incoherent. 
What is particularly striking here is that the change is not gradual but there appears to be two distinct regions
of the Fermi surface. This distinction between nodal and antinodal excitations is also found in underdoped BSSCO,\cite{ZX_old} and 
coincides with the abrupt onset of the pseudogap in the antinodal region.\cite{Kanigel} Recently, it was found from ARPES studies on La$_{2-x}$Ba$_x$CuO$_4$ 
that a d-wave like pseudogap state with nodal quasiparticles coexists with a stripe ordered state at $x=1/8$.\cite{Valla}
That study also shows that as a function of doping the pseudogap is actually maximized in the $x=1/8$ stripe ordered and non-superconducting state 
indicating an intimate relationship, at least in this material, between stripes and low-energy electronic spectral properties.

Here we present a model which is a caricature of a doped antiferromagnet with stripe correlations 
for which we find that the electronic spectral weight in the two 
regions of the 
Brillouin zone derive from spatially separated regions. The spectral weight in the nodal region comes from lightly hole-doped and 
antiferromagnetic ``spin-stripes'' whereas the antinodal spectral weight comes from more heavily doped ``charge-stripes'' which are antiphase domain
walls of the antiferromagnetic order.\cite{ovchinnikova} 
What is particularly striking is that the bandwidth transverse to the stripe extension is very small for the charge-stripe
states while for the spin-stripe states it is broad.
In the presence of disorder, intrinsic or in the form of a finite correlation length of the stripe order, the antinodal states are readily 
localized on individual charge stripes whereas the nodal states have a significantly longer localization length, remaining essentially two dimensional.  
Given the quasi one-dimensional nature of the antinodal states we can expect these to be very sensitive to the effects of electron-electron interactions. 
In fact, the low energy limit of an interacting one-dimensional electron gas is a Luttinger liquid for which the single hole spectral function is incoherent due to 
spin-charge separation and which for sufficiently strong interactions has no resemblance of a quasiparticle.\cite{Erica,Dror}
In addition there are many suggestions for the mechanism by which a one-dimensional electron gas may acquire a spin gap through interactions with the 
environment\cite{Steve_review}, thus providing a very natural connection with the antinodal pseudogap.

Several scenarios for the dichotomy already exist
including decoherence by scattering of antinodal quasiparticles with incommensurate spin fluctuations\cite{DH_Lee}
and explicit calculations on the Hubbard model using dynamical mean field theory supplemented by external spin or charge density fluctuations or extended
to larger clusters of sites.\cite{DMFT} 
These calculations are certainly more sophisticated than the present analysis which is a Hartree-Fock calculation that ignores any correlations beyond the assumed
stripe order. However, when supplemented by known results for a quasi one-dimensional electron gas the present work can provide a 
framework for understanding the nodal-antinodal dichotomy within a stripe scenario as outlined above.  

An additional intriguing feature is an  
even/odd effect of the charge periodicity for bond-centered stripes where for certain realizations of the model, ordered systems with even periodicity  
of the charge have a gap in the nodal region which is absent for odd periodicity. 
In a more realistic system with fluctuating stripes we speculate that this may result in 
a lower fraction of spectral weight in the nodal region for systems with even periodicity than systems with odd periodicity, thus  
possibly related to
the observation of ``magic doping fractions'' in LSCO with enhanced or suppressed low-temperature resistivity as a function of doping as well as the 
fact that the $1/8$ doped samples with charge stripe period 4 are especially prone to localization at low temperatures or by impurity 
doping.\cite{Ando_magic,AndoII}

\section{The model} 
\begin{figure}
\includegraphics[width=8cm]{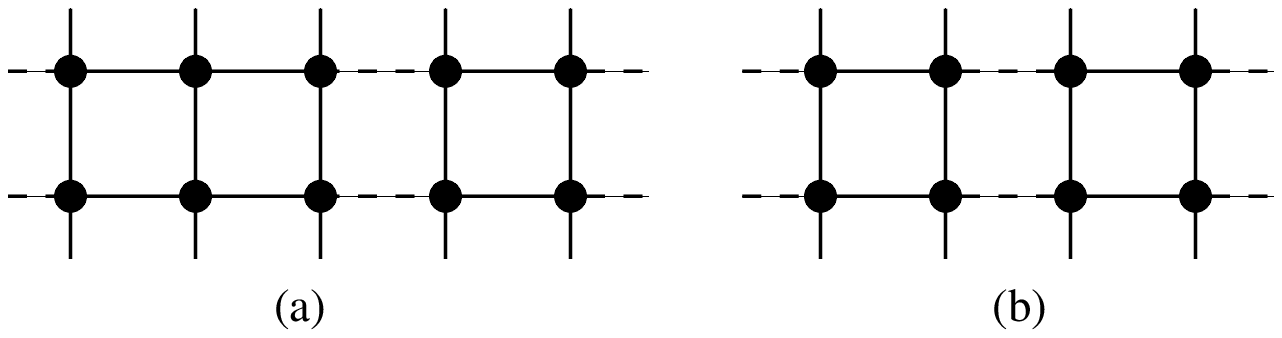}
\caption{\label{fig:array} Sketch of the (a) period-5 and (b) period-4 bond-centered stripe arrays, as defined in the text, 
consisting of the two and three leg Hubbard ladders given by the solid lines. 
Dashed lines indicate the coupling between ladders in the form of tunable hopping.}
\end{figure}

We consider an array
of $t-t'-U$ Hubbard ladders of varying width with on-site repulsion $U$ and nearest and next-nearest neighbor hopping $t$ and $t'$. A ladder is 
coupled to its neighboring ladders by tunable hopping $\lambda t$ and $\lambda t'$, where 
$0\leq\lambda\leq 1$. $\lambda$ is an ad hoc parameter which tunes between a quasi-one-dimensional problem at $\lambda=0$ and a fully two-dimensional problem 
(2D Hubbard) at $\lambda=1$. We are of course mostly interested in the $\lambda=1$ model, but when studying stripe solutions will find it 
quite illuminating to be able to adiabatically change the system from quasi-1D to 2D. 
Units are set in terms of $t=1$ and for the subsequent calculations we will take $t'=-.4$ and $U=4$. 
The value of $U$ is chosen such as to correspond to an optimal stripe filling of $0.5$ holes per unit length in a calculation with variable 
filling.\cite{later} We write coordinates $\bm{r}=(i,x)$, where $i$ is the chain index (horizontal coordinate in Fig.\ref{fig:array}) and
$x$ runs along the chains (vertical in Fig.\ref{fig:array}), and define electron creation and annihilation operators $\cdop{\bm{r}\sigma}$ and
$\cop{\bm{r}\sigma}$ with z-component of spin $\sigma=\up,\down$. The Hubbard interaction is $Un_{\bm{r}\up}n_{\bm{r}\down}$ where 
$n_{\bm{r}\sigma}=\cdop{\bm{r}\sigma}\cop{\bm{r}\sigma}$ is the number operator.

Clearly, we can not solve this problem of coupled Hubbard ladders, our ambition is only to formulate an effective theory for the distribution of 
electronic spectral weight in a stripe ordered antiferromagnet. For this purpose we will only consider spin and charge density order consistent with stripes.
We assume antiferromagnetic order and uniform density along the ladders and transverse spin and charge density waves   
with $\langle \vec{S}_{ix}\rangle=(-1)^x\hat{z}M_{i}$ ($-\half\leq M_i\leq\half$) and $\langle n_{ix} \rangle=\bar{n}+\delta n_i$ where $\delta n_i$ is 
the deviation from the mean density $\bar{n}=1-p$ with $p$ the hole doping.   
We solve for 
the magnetization and density on the ladders self-consistently using
$\langle n_{\bm{r}\up}n_{\bm{r}\down}\rangle=\frac{1}{4}\langle n_{\bm{r}}\rangle^2-\langle\vec{S}_{\bm{r}}\rangle^2$, and expand to linear order in the 
fluctuations around $\langle \vec{S}_{ix}\rangle$ and $\langle n_{ix} \rangle$ to derive an effective 
Hamiltonian $H_{\text{eff}}=H_t+H_I$. The tight-binding piece of the Hamiltonian reads
\begin{eqnarray}
&H_t&\!\!=-t\!\!\!\!\!\!\sum_{\langle \bm{r}\bm{r'}\rangle_{\text{intra}}}\!\!\!\!(\cdop{\bm{r}\sigma}\cop{\bm{r'}\sigma}+\mathbf{h.c.})
-t'\!\!\!\!\!\!\!\!\sum_{\langle\langle \bm{r}\bm{r'}\rangle\rangle_{\text{intra}}}\!\!\!\!\!\!(\cdop{\bm{r}\sigma}\cop{\bm{r'}\sigma}+\mathbf{h.c.})\nonumber\\
&-&\!\!\!\!\!\!\lambda t\!\!\!\!\!\!\sum_{\langle \bm{r}\bm{r'}\rangle_{\text{inter}}}\!\!\!\!(\cdop{\bm{r}\sigma}\cop{\bm{r'}\sigma}+\mathbf{h.c.})
-\lambda t'\!\!\!\!\!\!\!\!\sum_{\langle\langle \bm{r}\bm{r'}\rangle\rangle_{\text{inter}}}\!\!\!\!\!\!(\cdop{\bm{r}\sigma}\cop{\bm{r'}\sigma}+\mathbf{h.c.})
\end{eqnarray}  
where $\langle \bm{r}\bm{r'}\rangle$/$\langle\langle \bm{r}\bm{r'}\rangle\rangle$ indicates nearest/next nearest neighbor sites 
and  ``intra''/``inter'' indicate sites which are on the same/different ladders. The part of the effective 
Hamiltonian deriving from interactions reads
\begin{eqnarray}
H_I&=&\sum_{ix}(-1)^xm_i(n_{ix\up}-n_{ix\down})-\sum_{ix}m_iM_i\nonumber\\
&&+\sum_{ix}V_i(n_{ix\up}+n_{ix\down})-\frac{1}{2}\sum_{ix}V_i\delta n_i
\end{eqnarray}
where $m_i=-UM_i$ and $V_i=U\delta n_i/2$ are the effective staggered magnetization and potential due to the spin and charge density waves respectively. 

We should note that in the present work we only consider static stripes, 
whereas in the cuprates static stripes only occur under special circumstances.\cite{Tranquada_review} 
The use of a static stripe configuration to study electronic 
spectral properties in a system where there may only be fluctuating stripes thus assumes that the stripe fluctuations are slow compared to the spectral frequencies 
studied. 
Inelastic neutron scattering suggests that the 
frequency of stripe fluctuations is on the order of a few to around 50 meV ($10^{12}-10^{13}$Hz) depending on the material which is in fact comparable to the 
pseudogap energy, thus at least not ruling out this ``frozen'' stripe assumption.\cite{spin_gap}

\section{Band structure} 
We now proceed to find self-consistent solutions to this effective model, 
considering systems made up of alternating two-leg ladders and 
ladders of width $N$, denoting such an array by the period $N+2$. 
Here we explore only solutions where the two-leg ladders have $(0,\pi)$ (anti-phase) spin order and 
the $N$-leg ladder has a dominant AF $(\pi,\pi)$ component, i.e. bond-centered stripes. 
We call the two-leg ladder, which will self-consistently contain most of the doped holes,
``charge-stripe'' and the width $N$ ladder ``spin-stripe'',
and take the doping to be $0.5$ holes per unit length per charge-stripe (i.e. per $N+2$ charge period). 
These are solutions which obey the canonical 
relation between magnetic ``incommensurability'' and doping.\cite{Tranquada_review}

\begin{figure}
\includegraphics[width=8cm]{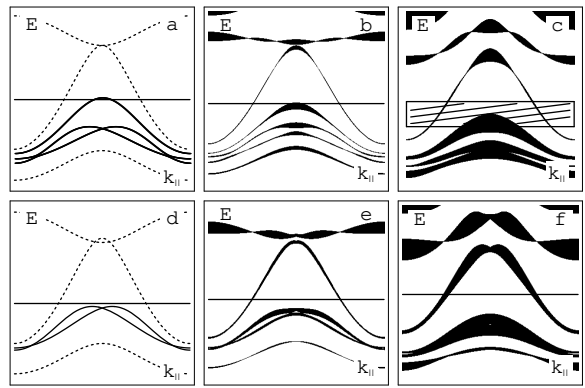}
\caption{\label{fig:bs} Band structure of the charge period 5 (top) and 4 (bottom) bond-centered stripe arrays showing the evolution with interstripe
hopping for $\lambda=0$ (left), $\lambda=0.5$ (center) and 
$\lambda=1$ (right). In a and d the dashed lines correspond to ``charge-stripe'' states and the solid lines to ``spin-stripe'' states as defined in the text.  
The energy window is $-3$ to $0.5$ (units of $t$) and the momentum $k_\parallel$ along the stripes ranges from $0$ to $\pi$ 
(units of inverse lattice spacing). 
The Fermi energy which corresponds to $10\%$ (a-c) and $12.5\%$ (d-f) hole doping is given by 
the horizontal lines. The dashed region in c shows the integration window for the ``Fermi surface'' in Fig. \ref{fig:FS}.}
\end{figure}

\begin{figure}
\includegraphics[width=8cm]{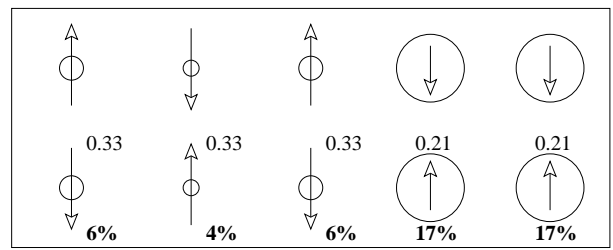}
\caption{\label{fig:densities} Fraction of one hole (bold) and direction and magnitude of the magnetization for the period-5 array with 
uniform hopping ($\lambda=1$). 
The magnetic unit cell pictured contains one doped hole for an overall doping of 10$\%$.}
\end{figure}

Our main results will follow from studying the period-4 and period-5 arrays which correspond to $12.5\%$ and $10\%$ doping respectively.   
The 4-array has an $8\times 2$ magnetic unit cell with magnetization $(-1)^x(M_e,-M_e,M_s,M_s,-M_e,M_e,-M_s,-M_s)$ and density variation 
$(\delta,\delta,-\delta,-\delta,\ldots)$
whereas the 5-array has a $5\times 2$
unit cell with $(-1)^x(M_e,-M_c,M_e,-M_s,-M_s)$ and 
$(\frac{2}{3}\delta-\frac{1}{2}\delta_c,\frac{2}{3}\delta+\delta_c,\frac{2}{3}\delta-\frac{1}{2}\delta_c,-\delta,-\delta)$ (see Fig. \ref{fig:densities}), where
all of the $M$ and $\delta$ parameters are solved for self-consistently. 
Fig \ref{fig:bs} shows the evolution with $\lambda$ of the band structure for the longitudinal momentum $k_\parallel$. 
The antiferromagnetic scattering along the stripes folds the bands
around $k_\parallel=\pi/2$, opening gaps between an upper and lower branch and 
the interladder hopping successively broaden the bands. 
For the one-dimensional band structure (Fig \ref{fig:bs} a and d) the (dashed) charge-stripe bands are roughly the same for both arrays with the
lower branch of the antibonding band of the two-leg ladder crossing the Fermi level around $\pi/4$ and $3\pi/4$. 
The spin-stripe bands on the
other hand are distinct as they derive from 3-leg and 2-leg ladders respectively. 
  
\section{Nodal-antinodal dichotomy} 
\begin{figure}
\includegraphics[width=8cm]{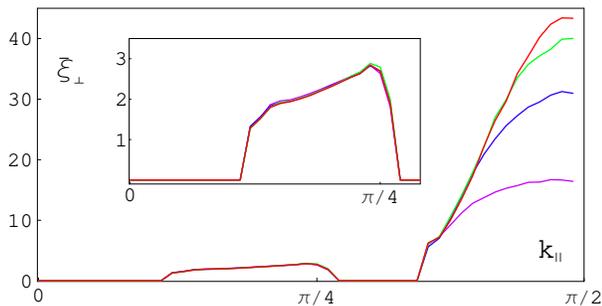}
\caption{\label{fig:loc_length} (color online) Mean localization length versus momenta along the stripes of all states in a $0.5$ window below the Fermi level 
for a period-5 
array with weak ($a=0.1$) one-dimensional quenched potential disorder. Inset shows only the ``charge-stripe'' states with $k_\parallel\alt\pi/4$.
Results are averaged over at least 10 runs with transverse system size $L_\perp=$100, 300, 600, and 900 corresponding to increasing $\bar{\xi}_\perp$. 
($\bar{\xi}_\perp(k_\parallel)=0$ means that there are no states in the integration window at that $k_\parallel$.)
}
\end{figure}

Now for the main observation of the paper; the active charge-stripe band stays very slim even when the hopping is fully 
two-dimensional (Fig. \ref{fig:bs}, c and f), implying a very small dispersion in the transverse direction, i.e. essentially one-dimensional. 
This is in sharp contrast to the behavior of the spectral weight below the Fermi level at
momenta around $\pi/2$ where the fatter distribution implies a significant dispersion with the transverse momentum. 
To make this distinction clearer we introduce disorder. 
In order to be able to 
work with large system sizes we will use one dimensional quenched disorder in the form of a weak random potential on each chain such that
$H_{\text{disorder}}=a\sum_{ix}\varphi_i n_{ix}$ where $-1\leq\varphi_i\leq 1$ is a random variable.  
We define the transverse
participation ratio of a state $\psi_{ix}$ as $P_\perp=\sum_i (\sum_x |\psi_{ix}|^2)^2 \approx \frac{1}{2}\xi^{-1}_\perp$ 
with $\xi_\perp$ being the transverse localization length. Fig. \ref{fig:loc_length} shows the mean of the localization lengths 
of the states near the Fermi level for the period-5 array with uniform hopping for weak disorder $a=.1$ and using a flat distribution for $\varphi$. 
(Here we do not solve the problem self-consistently, but simply use the values 
for the staggered magnetization and inhomogeneous potential of the ordered array.)    
We find that the localization length of the charge-stripe states ($k_\parallel\alt\pi/4$) is significantly shorter than that
of the spin-stripe states ($k_\parallel\agt\pi/4$) as expected from the difference in transverse bandwidth. The localization length of the
charge-stripe states is of the order of the stripe spacing, thus essentially localized on individual charge-stripes. 
The localization length of the spin-stripe states
is significantly larger, these states are essentially two-dimensional on the scale of the stripe period.     
We have checked that this aspect of the problem with narrow charge-stripe bands and broad spin-stripe bands comes out the same also 
if we consider
a system of site-centered stripes modeled as a single chain instead of the bond-centered two-leg ladder. In fact, we expect that this is a generic property 
which follows 
from the antiphase magnetic structure containing in-gap ``impurity'' states of the surrounding antiferromagnet.\cite{stripe_spectral}

\begin{figure}
\includegraphics[width=8cm]{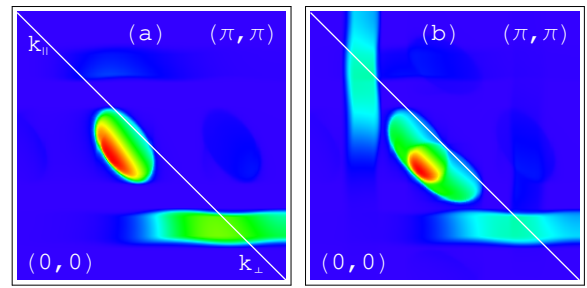}
\caption{\label{fig:FS} (color online) ``Fermi surface'' in the first quadrant of the Brillouin zone of the period-5 array with uniform hopping $\lambda=1$ and
weak disorder ($a=.1$) as in Fig. \ref{fig:loc_length}. 
The spectral weight is integrated over a $0.5$ window 
below the Fermi level. (a) is for stripes along one direction ($\parallel$) and (b) is symmetrized with respect to the stripe direction.
(sampled over 10 runs with system size $L_\perp=500$).}
\end{figure}

Fig. \ref{fig:FS}a shows the low-energy spectral weight integrated over a small ($0.5$) window (to account for real-world broadening) below the Fermi level 
of the period-5 array, corresponding roughly 
to the dashed region in Fig. \ref{fig:bs}c, for the same weak disorder realization ($a=.1$) as above.  
As discussed in earlier work the bulk of spectral weight remains close to the undoped tight-binding 
Fermi surface.\cite{stripe_spectral,stairs} The fact that the Fermi surface is broad is a combination of the scattering by the 
stripe potential and the finite integration window. In addition, with disorder the states are localized in the transverse stripe direction and thus broad in the 
transverse momentum ($k_\perp$), with the width proportional to the inverse localization length.  
There are two disjoint patches of spectral weight which we can readily identify as deriving from the different stripe 
bands, where the weight in the nodal region, near  $(\pi/2,\pi/2)$,
comes from the spin-stripes whereas the spectral weight in the antinodal region, near $(\pi,0)$, 
comes from the charge stripes. Fig. \ref{fig:FS}b is symmetrized with respect to the stripe direction showing a qualitative agreement with the 
ARPES data\cite{ZX_nodal} with the characteristic straight Fermi surface sections. 
Considering now the
results for the localization length of the same system presented in Fig. \ref{fig:loc_length}, we can thus make the rather profound statement that, at least for 
this model system, 
the nodal states are two-dimensional while the antinodal states are quasi-one-dimensional. 

We should emphasize that the agreement between 
Fig. \ref{fig:FS}b and the various ARPES measurements of the Fermi surface is very rough as in the ARPES
data the antinodal weight at the Fermi energy is suppressed by the pseudogap and most of the spectral weight is in the nodal region ``Fermi arc''.\cite{ZX_Fermiarc} 
In this work we do not consider any correlation effects apart from the stripe order itself. 
However, as discussed earlier there are several scenarios, including the ``spin-gap proximity effect'' by Emery, Kivelson, and Zachar, 
for how a one-dimensional electron gas may acquire a spin gap through interactions with the 
environment.\cite{Zachar,Steve_review} Based on this a natural extension of the present work is to assume that the quasi-one-dimensional hole-rich charge-stripes
which are connected to the antinodal spectral weight 
are akin to spin-gapped Luttinger liquids, or Luther-Emery liquids. 
For a Luther-Emery liquid the single hole spectral function is incoherent due to 
spin-charge separation with suppressed spectral weight at the Fermi energy due to the spin gap.\cite{Dror} 
The spin gap in this scenario is thus effectively the pseudogap which suppresses the antinodal spectral weight at the Fermi energy.


%

\section{Magic filling fractions} 
\begin{figure}[t]
\includegraphics[width=8cm]{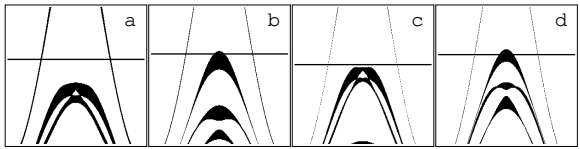}
\caption{\label{fig:bs2345} Truncated band structure of the charge period (a) 4 (b) 5 (c) 6 and (d) 7 bond-centered stripe arrays, 
with Fermi energy (horizontal line) 
corresponding to $1/8$, $1/10$, $1/12$, and $1/14$ doping respectively, at intermediate interladder coupling 
$\lambda=0.5$. (Energy window -2 to -1 in units of $t$, else as in Fig.\ref{fig:bs})}
\end{figure}

For bond-centered stripes as envisioned here, there is an interesting even-odd effect of the stripe periodicity. 
Spin-stripes of odd width have a high-energy
lower band with transverse momentum $k_\perp =\pi/2$ and a maximum at $k_\parallel=\pi/2$ which is absent for even width spin-stripes.
Fig. \ref{fig:bs2345} shows the band structures at intermediate, $\lambda=0.5$, 
coupling of period-$N+2$ arrays with spin stripe width $N=$2 to 5 where this distinction is seen. (For the $N=4$ and 5 systems we have 
assumed a uniform magnetization and density on the spin-stripes, which we have checked is a good approximation.) 

We may speculate that a signature of this even-odd feature survives in the real materials where the actual gap in the nodal region of the even period 
systems is destroyed by stripe fluctuations or interaction broadening and instead manifested as a smaller fraction of nodal spectral weight.
Assuming that the main contribution to the low-temperature conductivity comes from the ``nodal metal''\cite{nodal_metal} 
it is thus tempting to identify this with the observation of ``magic doping fractions'' exhibited in the low-temperature in-plane resistivity of
LSCO where there is a non-monotonous variation with doping.\cite{Ando_magic} 
It was suggested that the magic fractions would be the effect of a checkerboard type order and not consistent with stripe order\cite{Ando_magic}, however, 
the overall modulation of the low temperature resistivity
seem to fit quite well with a suppressed conductivity at doping $1/8$, $1/12$ and $1/16$ and enhanced conductivity at $1/10$ and $1/14$. In addition 
the $1/8$ doped
system (Fig. \ref{fig:bs2345}a) would have a particularly large fraction of antinodal spectral weight which being quasi-one-dimensional is 
very sensitive to disorder, possibly related to the anomalous behavior at this doping.\cite{AndoII}

\section{Conclusions} We study a model of a stripe ordered doped antiferromagnet which can be tuned from quasi one-dimensional to two-dimensional, 
solving self-consistently for the magnitude of the spin and charge order.
We find that the electronic spectral weight in the nodal and antinodal regions of the 
Brillouin zone derive from spatially separated regions with the nodal weight coming from lightly hole-doped and 
antiferromagnetic ``spin-stripes'' whereas the antinodal spectral weight comes from more heavily doped ``charge-stripes''. 
The main result is that the bandwidth transverse to the stripe extension is very small for the charge-stripe
states while for the spin-stripe states it is considerably wider.
Thus, in the presence of disorder, intrinsic or in the form of a finite correlation length of the stripe order, the antinodal states are 
localized on individual charge stripes with a localization length which is typically shorter than the inter-stripe spacing whereas the nodal states have a 
significantly longer localization length, remaining essentially two dimensional. 
In combination with known results for the interacting one-dimensional electron gas these results provide a framework for understanding the nodal-antinodal 
dichotomy of the low-energy electronic spectral properties of the underdoped cuprates: The antinodal spectral weight is quasi one-dimensional and incoherent due
to electron fractionalization with a spin gap manifesting itself as the pseudogap whereas the 
nodal weight, which is less affected by the stripe order and thus effectively two-dimensional, is more normal, possibly Fermi-liquid like with electronic 
quasiparticle excitations.

We also identify an even-odd effect of the charge-stripe periodicity for certain realizations of the model 
in which for a system with ordered bond-centered stripes even period stripes have a spectral gap 
in the nodal region which is absent for odd period systems. Although this topic needs to be studied in more detail we speculate that this effect may be 
related to the non-monotonous variations with doping of the low-temperature resistivity of LSCO.

I would like to thank Dror Orgad, Shirit Baruch and Marina Ovchinnikova for valuable discussions and John Tranquada for helpful comments.


\begin{thebibliography}{12}
\bibitem{spin_gap} Here we interpret the lower edge (spin gap) of the incommensurate branches as the stripe fluctuation frequency and the pseudopgap energy 
is the typically energy scale over which we study the low-energy spectral weight.
\end{thebibliography}
\end{document}